\documentclass[reqno]{article}

\usepackage{graphicx}

\usepackage{hyperref}
\hypersetup{
    colorlinks = true,
    urlcolor   = blue,
    citecolor  = black,
}
\usepackage[dvipsnames]{xcolor}
\usepackage{amssymb,amsthm}
\usepackage[utf8]{inputenc}
\usepackage[numbers]{natbib}
\usepackage{xcolor}
\usepackage{hyperref}
\usepackage{amsxtra}
\usepackage{mathtools}
\usepackage{newtxtext}
\usepackage{newtxmath}
\mathtoolsset{showonlyrefs=true}
\usepackage{booktabs}
\usepackage[margin=1em]{subfig}
\usepackage{todonotes}
\usepackage{layouts}

\newcommand{\RomanNumeralCaps}[1]
\linenumbers

\newtheorem{theorem}{Theorem}[section]
\newtheorem{proposition}{Proposition}[section]

\newtheorem{remark}{Remark}[section]

\newcommand{\Rr}{\mathbb{R}}

\newcommand{\Ss}{\mathbb{S}}

\newcommand{\SU}{\mathfrak{su}}
\newcommand{\UU}{\mathfrak{u}}

\newcommand{\Tr}{\mbox{Tr}}
\newcommand{\Stab}{\mbox{stab}}
\newcommand{\norm}[1]{\lVert #1 \rVert}

\newtheorem{ass}{Assumption}

\title{Canonical Scale Separation in 2D Incompressible Hydrodynamics}

\author{Klas Modin\footnote{Chalmers University of Technology and University of Gothenburg, Gothenburg, Sweden - klas.modin@chalmers.se},
 \and Milo Viviani\footnote{Scuola Normale Superiore di Pisa, Pisa, Italy}}

\begin{document}

\maketitle

\begin{abstract}
A two-dimensional inviscid incompressible fluid is governed by simple rules. Yet, to characterise its long-time behaviour is a knotty problem. The fluid evolves according to Euler's equations: a non-linear Hamiltonian system with infinitely many conservation laws. In both experiments and numerical simulations, coherent vortex structures, or blobs, emerge after an initial stage. These formations dominate the large-scale dynamics, but small scales also persist.  Kraichnan describes in his classical work a forward cascade of enstrophy into smaller scales, and a backward cascade of energy into larger scales. Previous attempts to model Kraichnan's double cascade use filtering techniques that enforce separation from the outset. Here we show that Euler's equations posses an intrinsic, canonical splitting of the vorticity function. The splitting is remarkable in four ways: (i) it is defined solely via the Poisson bracket and the Hamiltonian, (ii) it characterises steady flows, (iii) without imposition it yields a separation of scales, enabling the dynamics behind Kraichnan's qualitative description, and (iv) it accounts for the ``broken line'' in the power law for the energy spectrum, observed in both experiments and numerical simulations. The splitting originates from Zeitlin's truncated model of Euler's equations in combination with a standard quantum-tool: the spectral decomposition of Hermitian matrices.  In addition to theoretical insight, the scale separation dynamics could be used for stochastic model reduction, where small scales are modelled by multiplicative noise.
\end{abstract}

%


\section{Introduction}
Two-dimensional turbulence is the study of incompressible hydrodynamics at large (including infinite) Reynolds numbers.
It is a vibrant field of both mathematics and physics that began with \citet{Eu1757}, who derived the basic equations of motion. 
Turbulent flows in two space dimensions do not exist as classical fluids in nature.
Rather, they constitute basic models of intermediate-scale flows in ``almost'' two-dimensional (thin) domains~\citep{MaBe2002,MaWa2006,Cu2006,Pe2013,Do2013,Ze2018}.

\begin{figure*}
  \setlength{\unitlength}{\textwidth}
  \begin{picture}(1,0.185)
    \put(-0.03,-0.02){\includegraphics[width=1.05\textwidth]{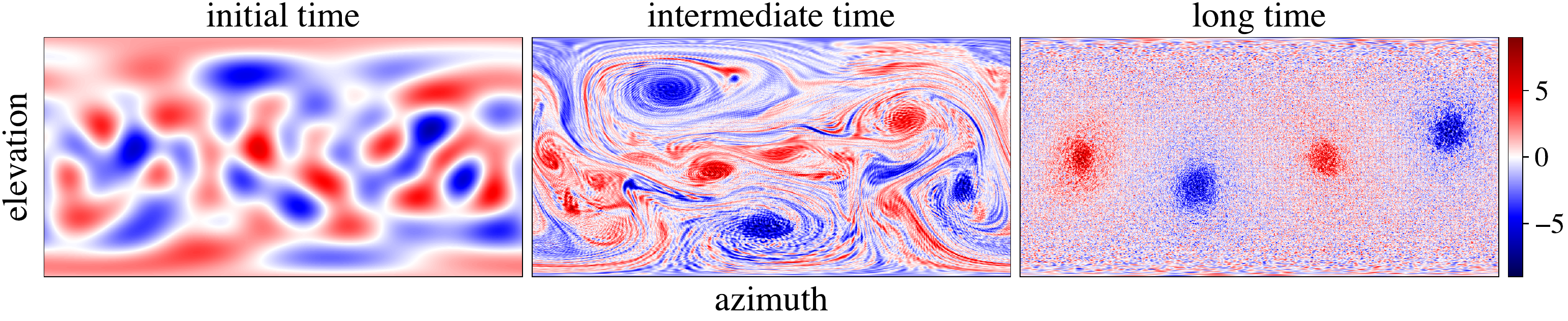}} 
  \end{picture}
  \caption{Evolution of vorticity for Euler's equations on the sphere. 
  Vorticity regions of equal sign undergo merging to form stable, interacting vortex condensates.
  }\label{fig:evolution1}
\end{figure*}

The conditions of 2D turbulence can be emulated in experiments.
One setup is a soap film flowing rapidly through a fine comb \citep{Co1984}. 
Another is a conducting fluid confined to a thin layer and driven into turbulence by a temporally varying magnetic field \citep{So1988}.
When such a ``quasi-2D'' flow is released it self-organises into blob-like condensates.
The progression is depicted in \autoref{fig:evolution1} for a spherical domain.
Heuristically, the mechanism is driven by merging of equally signed vorticity regions. 
This large scale fusion is balanced by fine scale dissipation. 
In many ways 2D turbulence is propelled by the quest to understand the resulting scale separation.

To make theoretical progress, \citet{On1949} applied statistical mechanics to a large but finite number of \emph{point vortices}.
They are weak (distributional) solutions of Euler's equations where vorticity is a sum of weighted Dirac pulses.
Onsager realised that a fixed number of positive and negative point vortices, confined to a bounded domain, can have energies from $-\infty$ to $\infty$.
The phase volume function therefore has an inflection point at some finite energy.
At this energy the thermodynamical temperature is zero.
If the energy exceeds this point, so the temperature is negative, then equally signed vortices should cluster according to thermodynamics. 
This theory of \emph{statistical hydrodynamics} 
is a prominent, although lesser known, part of Onsager's legacy~\citep{EySr2006}.
On the mathematical side, \citet{CaLiMaPu1992,CaLiMaPu1995} and \citet{Ki1993} gave rigorous results on clustering of point vortices in the negative temperature regime.
Their work has fostered much theoretical progress (see \citet{MaPu2012} and references therein).
On the experimental side, seventy years after Onsager presented his theory, the conditions of negative temperature point vortex dynamics were experimentally realised in planar Bose-Einstein condensates~\citep{GaReYuBr2019,JoGrStBi2019}.
As predicted, persisting vortex clusters emerge.

Onsager's theory cannot be applied to continuous vorticity fields (corresponding to smooth solutions).
Consequently, it is natural to look for a statistical theory of continua.
One approach is to expand the vorticity field in a Fourier series and then truncate it~\citep{Kr1975}.
The truncated, finite-dimensional system preserves phase volume and quadratic invariants, but not higher order invariants (Casimir functions).
To account also for those invariants, another approach is to maximise the entropy of a probability distribution of coarse-grained vorticity fields~\citep{Lyn1967,Mi1990,RoSo1991}.

All theories based on statistical mechanics assume \emph{ergodic dynamics}.
Rigorous results on ergodicity are available for 2D Navier-Stokes on a doubly periodic domain (flat torus) with added regular-in-space noise proportional to the square root of the viscosity $\nu$.
In this setting there exists a unique stationary measure $\mu_\nu$~\citep{Ku2012,KuSh2017}.
As $\nu\to 0$ one obtains a stationary measure $\mu_0$ for the 2D Euler equations, but it is not expected to be unique~\citep{KuSh2017}.

Statistical mechanics is not the only approach. 
Another possibility is to study energy and enstrophy spectra.
The inspiration comes from Kolmogorov's (\citeyear{Ko1941}) theory of 3D turbulence.
Notably, \citet{Kr1967} argued that viscous 2D fluids in \emph{forced equilibrium}, where energy at an intermediate scale is fed into the system, exhibit a forward cascade of enstrophy into fine scales and a simultaneous backward cascade of energy into large scales.
Direct numerical simulations typically support Kraichnan's theory \citep[and references therein]{XiWaChEy2009}.

For two-dimensional systems with no energy dissipation at the large scale (so that vortex condensation occurs), numerical simulations develop a ``broken line'' energy spectrum with a slope of $k^{-3}$ at the large scale (where $k$ is the wave-number) and then a swift switch at an intermediate scale to a slope between $k^{-5/3}$ and $k^{-1}$~\citep{BoEc2012}.
An approximate 
broken line energy spectrum is also observed in zonal and meridional wind measurements on Earth over the scales 3--10,000~km~\citep{NaGaJa1984}.

To better understand characteristic energy spectra it is natural to impose a splitting of the vorticity field $\omega = \omega_s + \omega_r$ into a large-scale component $\omega_s$ and a small-scale component $\omega_r$.
A wavelet-based vorticity splitting is proposed by \citet{FaScKe1999} and applied to numerical simulation on the doubly periodic square (where condensation occurs) by \citet{ChCoKoLe2007}.
The results \citep[fig.~1f]{ChCoKoLe2007} show an energy spectrum slope of $k^{-3}$ for the large scale component 
and of $k^{-1}$ for the intermediate-to-small scale component.
It is a powerful technique to analyse energy spectra, but it depends on a choice of wavelet basis and on parameters identifying the different scales.
Therefore the method cannot give insights on the \emph{mechanisms} behind vortex condensation or broken line energy spectra.

In this paper we give a new, canonical decomposition of vorticity.
With ``canonical'' we mean parameter-free, determined solely in terms of the data for the two-dimensional Euler equations: the Poisson bracket and the Hamiltonian function.
The decomposition has the following properties:
\begin{enumerate}
  \item The vorticity field $\omega = \omega_s + \omega_r$ is a steady state if and only if $\omega_r = 0$.
  \item Under numerical simulation of Euler's equations, $\omega_s$ and $\omega_r$ evolve into a separation of scales.
  The component $\omega_s$ traps large-scale condensates whereas the component $\omega_r$ captures small-scale fluctuations.
  \item After a short transient time, the energy spectrum slope of $\omega_s$ is about $k^{-3}$ and of $\omega_r$ is between $k^{-5/3}$ and $k^{-1}$.
  \item Over time, the component $\omega_r$ displays an average enstrophy increase (quantifying the forward enstrophy cascade) and an average energy decrease (quantifying the backward energy cascade).
\end{enumerate}

The coupled equations governing the dynamics of $\omega_s$ and $\omega_r$ embody a new line of attack for 2D turbulence.
Our standpoint is that a detailed study of these equations may explain the mechanisms behind vortex condensation and broken line energy spectra, or at least give deep insights.
The numerical simulations we present suggest so.

\subsection{Two-dimensional Euler equations}

Our starting point is Euler's equations for an inviscid, incompressible fluid on a two-dimensional closed surface.
Throughout the paper we take the surface to be the unit sphere $\Ss^2\subset \mathbb{R}^3$. 
It makes our arguments more explicit and enables numerics.
Most concepts are readily transferable to arbitrary closed surfaces (in particular to the flat torus, which is the most studied example in the literature albeit less relevant than the sphere in applications).

In vorticity formulation, Euler's equations on $\Ss^2$ are
\begin{equation}\label{eq:euler_eqn_1}
\dot{\omega} =\lbrace\psi,\omega\rbrace ,\qquad \Delta\psi = \omega,
\end{equation}
where $\{\cdot,\cdot\}$ is the Poisson bracket on $\Ss^2$, $\omega$ is the vorticity function of the fluid, related to the fluid velocity $\mathbf v$ via $\omega = \operatorname{curl}\mathbf{v}$, and $\psi$ is the stream function, related to the vorticity function via the Laplace-Beltrami operator $\Delta$.
Geometrically, the equations~\eqref{eq:euler_eqn_1} constitute an infinite-dimensional \emph{Lie--Poisson system} \citep[and references therein]{ArKh1998}.
The phase space consists of vorticity fields and is equipped with the following infinite-dimensional Poisson bracket
\[
  \prec \!F, G \!\succ\! (\omega) = \int_{\Ss^2} \left\{ \frac{\delta F}{\delta \omega}, \frac{\delta G}{\delta \omega}  \right\} \omega .
\]
The Hamiltonian (total energy) for the vorticity equation~\eqref{eq:euler_eqn_1} is
\[ H = -\dfrac{1}{2}\int_{\Ss^2}\psi\omega .\]
In addition to total energy, there is an infinite number of conservation laws:
total angular momentum
\[ \mathbf L = \int_{\Ss^2} \omega\mathbf{n}\, , \quad \text{$\textbf{n}$ unit normal on $\Ss^2$,}\]
and Casimir functions
\[C_f(\omega) = \int_{\Ss^2}f(\omega),\quad \text{for any smooth $f\colon \Rr\to \Rr.$}\]
These conservation laws are fundamental for the long-time behaviour.
In particular, the presence of infinitely many Casimir functions sets apart 2D from 3D fluids.

\subsection{Overview of the paper}


Zeitlin's truncated model of Euler's equations originate from the vorticity formulation~\eqref{eq:euler_eqn_1}.
In the spirit of quantization theory, the space of vorticity functions is replaced by the space $\UU(N)$ of skew-hermitian complex matrices, while the Poisson bracket is replaced by the matrix commutator.
The size $N$ of the matrices is the spatial discretization parameter, related to `Planck's constant' in quantum theory via $\hbar = 1/N$.
We present an overview of how functions and matrices are related in \autoref{sec:quant_EE}.

There are at least two advantages of Zeitlin's model. 
First, it yields a spatial discretization that preserve all the underlying geometry of the Euler equations: the Hamiltonian structure and the conservation laws.
Combined with a symplectic time-integration scheme one can obtain fully structure preserving numerical methods~\citep{MoVi2020}.
Second, complicated topological or geometrical properties of the Euler equations can be described in terms of standard tools from linear algebra and matrix Lie groups.
Indeed, our splitting naturally arise from the standard spectral decomposition of Hermitian matrices applied to the stream matrix.
The splitting has a precise geometric meaning in terms of Lie algebras, but also a dynamical interpretation as the steady and unsteady vorticity components.
The details are given in \autoref{sec:vort_split}.

Numerical experiments for the canonical splitting are presented and discussed in \autoref{sec:vort_numerics}.
We demonstrate that the components in the splitting convergence into a separation of scales.
In addition, they capture broken line energy spectra.

In \autoref{sec:vort_cont} we translate our results about Zeitlin's model to the continuous Euler equations.
All matrix concepts used for the canonical splitting have classical, fluid dynamical counterparts.
But to define them rigorously requires prudence.
It is essential to use a weak formulation.
The section sets the foundation for an $L^\infty$-based theory of canonical vorticity splitting, independent from Zeitlin's model.
It is the most mathematical part of the paper.
Even so, a heuristic explanation is straightforward.
Let us give it here.
For a smooth vorticity function $\omega$, the aim is to construct a splitting $\omega = \omega_s + \omega_r$.
%
%
Let $\gamma(\tau)$ be a closed level curve of $\psi$.
We define $\omega_s$, restricted to $\gamma$, to take the constant value
\begin{equation}
  \omega_s\big|_{\gamma} = \frac{1}{\operatorname{length}(\gamma)}\int_\gamma \omega\, d\tau .
\end{equation}
In other words, $\omega_s$ is obtained via averaging of $\omega$ along the level curves, or streamlines, of~$\psi$.
The mathematical difficulties arise where the level curves contain bifurcations.


\section{Background on Euler-Zeitlin equations}\label{sec:quant_EE}
In this section we give background on the ``consistent truncation'' of Euler's equations introduced by \citet{ze1, ze2}.
The model relies on quantization of the Poisson algebra $C^\infty(\Ss^2)$ of smooth functions on the sphere \citep{hopPhD,FaiZac1989}.
Recall that the Poisson bracket $\lbrace f, g\rbrace$ for $x\in\Ss^2\subset\mathbb{R}^3$ and $f,g\in C^\infty(\Ss^2)$ is the function given by
\[
\lbrace f,g\rbrace(x) = x \cdot (\nabla f\times\nabla g).
\]
In our context, quantization means to find a projection from smooth functions $C^\infty(\Ss^2)$ to complex skew-Hermitian matrices $\UU(N)$ such that the Poisson bracket under this map is approximated by the matrix commutator.

Quantization on the sphere is explicit.
Indeed, \citet{hopyau} gave an operator $\Delta_N$ on $\UU(N)$ that, up to the truncation index $N$, has the same spectrum as the Laplace-Beltrami operator on the sphere.
The eigenvectors of $\Delta_N$ are thereby quantized versions $T^N_{lm}$ of the spherical harmonics $Y_{lm}$.
By computing the eigenbasis $T^N_{lm}$, which is fast due to the diagonal structure of the Hoppe-Yau Laplacian, an explicit quantization scheme for the sphere is obtained.
It yields a projection map $\Pi_N\colon C^\infty(\Ss^2)\to \UU(N)$ relating functions to matrices via
\begin{equation}\label{eq:truncation}
  C^\infty(\Ss^2) \ni \omega = \sum_{l=0}^\infty\sum_{m=-l}^l\omega^{lm} Y_{lm} \mapsto 
  \sum_{l=0}^{N}\sum_{m=-l}^l \omega^{lm}T^N_{lm} = W \in \UU(N).
\end{equation}
For more details and explicit formulae we refer to \citet{hopyau,ze2,MoVi2020}.


The vorticity formulation~\eqref{eq:euler_eqn_1} uses only the Laplacian $\Delta$ and the Poisson bracket $\{\cdot,\cdot\}$.
With the quantized analogues $\Delta_N$ and $[\cdot,\cdot]$ we thereby obtain the \emph{Euler-Zeitlin equations}
\begin{equation}\label{eq:Euler_eqn_qant1}
\dot{W} =[P,W],\qquad \Delta_N P = W,
\end{equation}
where $W\in\SU(N)$ is the \emph{vorticity matrix} and $P\in\SU(N)$ is the \emph{stream matrix}.
The condition
\begin{equation}
  W \in \SU(N) = \{A\in \UU(N)\mid \operatorname{tr} A=0 \}
\end{equation}
correspond to vanishing total circulation $\int_{\Ss^2}\omega = 0$.

The Euler-Zeitlin equations~\eqref{eq:Euler_eqn_qant1} have been studied in various contexts, primarily on the flat torus~\citep{ze1, mcl, AbMa2003}, but more recently also on the sphere~\citep{ze2, MoVi2020}.
Their main feature is that they preserve the rich geometry in phase space of the original equations~\eqref{eq:euler_eqn_1}, namely the Lie-Poisson structure (see \citet{ModViv2019a,MoVi2020} for details).
In turn, this implies conservation of total energy $H(W) = \Tr(PW)/2$, (quantized) Casimirs $C_k(W) = \Tr(W^k)$, and angular momentum $\mathbf{L}=(L_x,L_y,L_z)$.
No conventional spatial discretization of the Euler equations preserve all these properties.


\begin{remark}
Although the spherical setting is more complicated to work with, 
the Euler-Zeitlin equations are actually more accurate on the sphere than on the torus.
There is a deep geometrical reason for this: quantization on the sphere exactly preserves the $\mathrm{SO}(3)$ symmetry, whereas the corresponding translational symmetry in the quantization of the torus is only approximately captured.
This difference is evident in the definition of the discrete Laplace operator on the sphere and the torus.
\end{remark}

In our previous work~\citep{ModViv2019a,MoVi2020} we develop a Lie--Poisson preserving numerical method for the Euler-Zeitlin equations on the sphere and we use it to study the long-time behaviour.
As reported also by \citet{dri}, the numerical results give strong evidence against the predictions of statistical mechanics theories, derived for the sphere by \citet{He2013}; see also \citet{BoVe2012}.
Rather, the results suggest the existence of near-integrable parts of phase space that act as barriers for the statistical predictions to be reached.
Those near integrable solutions take the form of interacting vortex blobs (3 or 4 depending on the total angular momentum), perfectly reflecting integrability results for Hamiltonian blob dynamics on the sphere \citep{MoVi2020b}.

During our work with Zeitlin's model a new point of view emerged. 
More than a spatial discretisation, the Euler-Zeitlin equations themselves provide tools for studying the long-time behaviour of 2D hydrodynamics.
Those tools include, in particular, Lie theory for $\UU(N)$, which is exceptionally well understood, for example from the point of quantum theory, representation theory, and linear algebra.
In particular, looking through the lens of Lie algebra theory, it is natural to split the vorticity matrix~$W$ by orthogonal projection onto the \emph{stabilizer} of the stream matrix~$P$; the underpinning of this paper.
By simulating the Euler-Zeitlin equations~\eqref{eq:Euler_eqn_qant1} using our Lie-Poisson integrator, and then for each output compute the canonical splitting, we see that it captures the dynamics of vortex condensation and scale separation, directly related to the theory of \citet{Kr1967} for an inverse energy cascade.




\section{Canonical splitting of the vorticity matrix}\label{sec:vort_split}
In this section we present and discuss canonical vorticity splitting for Zeitlin's model.
``Canonical'' means the splitting only depends on the Lie algebra structure (and, of course, on the vorticity matrix $W$ and the stream matrix $P$ governing the dynamics).
It does not require any \emph{ad hoc} choice of scale as previous methods do.
The separation of scales results from the dynamics.


Consider again the Euler-Zeitlin equations~\eqref{eq:Euler_eqn_qant1}.
Equip $\SU(N)$ with the Frobenius inner product. 
The canonical splitting of the vorticity matrix 
\begin{equation}\label{eq:canonical_splitting}
  W=W_s + W_r
\end{equation}
is obtained by taking $W_s$ to be the orthogonal projection of $W$ onto the \emph{stabilizer} of the stream matrix $P$
\begin{equation}
  \Stab_P = \lbrace A\in \SU(N)\mid [A,P]=0\rbrace .  
\end{equation}
If $P$ is generic, so all its eigenvalues are distinct, then $\Stab_P$ is equivalently given by
\begin{equation}
\Stab_P = \lbrace A\in \SU(N)\mid A,P\mbox{ simultaneously diagonalizable}\rbrace.
\end{equation}
In this case $W_s$ is obtained via the spectral decomposition: first find $E\in \mathrm{SU}(N)$ which diagonalizes $P$, i.e., $E^\dagger PE=\Lambda$, then set $\Pi_P:\SU(N)\rightarrow\Stab_P$ as
\[W_s:=\Pi_P(W)=E\mbox{diag}(E^\dagger WE)E^\dagger.\]

\begin{remark}
Relative to the splitting \eqref{eq:canonical_splitting}, the Euler-Zeitlin equations~\eqref{eq:Euler_eqn_qant1} can be written
\begin{equation*}
\dot{W} =[P,W_r].
\end{equation*}
Thus, $W = W_s + W_r$ is a steady solution (equilibrium) if and only if $W_r = 0$, so the dynamics is ``driven by'' the residual part $W_r$.
\end{remark}

\subsection{Dynamics of \texorpdfstring{$W_s$}{Ws} and \texorpdfstring{$W_r$}{Wr}}

For insight on the splitting~\eqref{eq:canonical_splitting} let us express the dynamics of the Euler-Zeitlin equations in the variables $W_s$ and $W_r$.
Consider first a general flow on $\SU(N)$ of the form
\begin{equation}\label{nonisoflow}
  \dot P = F(P),
\end{equation}
for some smooth vector field $F\colon\SU(N)\to\SU(N)$.
Let $E \in \mathrm{SU}(N)$ and $\Lambda \in \mathrm{diag}_N$ denote an eigenbasis and corresponding eigenvalues for $P$.
Given \eqref{nonisoflow} we first determine the evolution of $E$ and $\Lambda$.
The Lie algebra $\SU(N)$ is foliated in orbits (\emph{cf.}\ \citet{Ki2004}) given by
\[
\mathcal{O}_P = \lbrace UPU^\dagger \mid U\in \mathrm{SU}(N)\rbrace.
\] 
In the generic case (when all eigenvalues are distinct), the tangent space $T_P\mathcal{O}_P$ is spanned by $\{ \mathrm{i}e_k e_l^\dagger \}_{k\neq l}$, where $E=[e_1,\ldots,e_N]$ is an orthonormal eigenbasis of $P$.
The orthogonal directions $T_P\mathcal{O}_P^\bot = \mathrm{span}\{\mathrm{i}e_ke_k^\dagger\} = \Stab_P$ is the linear subspace of matrices in $\SU(N)$ sharing the same eigenbasis (they are simultaneously diagonalizable).
Thus the two projections
\begin{equation}
  \Pi_P\colon \SU(N)\to \mathrm{stab}_P  \hspace{.2cm}\text{and} \hspace{.2cm} \Pi_P^\bot \coloneqq \mathrm{Id} - \Pi_P \colon \SU(N)\to \Stab_P^\bot
\end{equation}
correspond to decomposition in the basis $\{\mathrm{i}e_ke_k^\dagger\}_k$ and $\{\mathrm{i}e_ke_l^\dagger\}_{k\neq l}$.
Notice, as expected, that neither $\Pi_P$ nor $\Pi_P^\bot$ depend on the eigenvalues of $P$, only on the eigenbasis.

We can now write equation~\eqref{nonisoflow} as 
\begin{equation}
  \dot P = \Pi_P F(P) + \Pi_P^\bot F(P) .
\end{equation}
The first part of the flow changes the eigenbasis but not the eigenvalues and vice versa.
The question is: what is the generator of $P\mapsto \Pi_P^\bot F(P)$? 
Since it is isospectral it should be of the form $P\mapsto [B(P),P]$ for some $B(P)\in\SU(N)$.
Let us denote $X=F(P)$. 
It is then straightforward to check that if all the eigenvalues $p_1,\ldots,p_N$ of $P$ are different, then
\begin{equation}
\begin{aligned}
  \Pi_P^\bot X &= \sum_{k\neq l} x_{kl}\mathrm{i} e_ke_l^\dagger
  = \sum_{k\neq l} (p_k-p_l)b_{kl}e_ke_l^\dagger\\
  &= \left[\sum_{k\neq l} b_{kl}e_ke_l^\dagger, P\right]
  = [B,P]
\end{aligned}
\end{equation}
where $x_{kl}$ are the components of $X$ in the basis $E$, and $b_{kl} \coloneqq x_{kl}/(p_k-p_l)$.
Thus, in the generic case we can construct the generator $B(P)$ from the eigenvalues $p_1,\ldots,p_N$ and the eigenbasis $e_1,\ldots,e_N$ of $P$.
This allows us to write equation~\eqref{nonisoflow} in terms of the eigenvalues and eigenbasis of $P$ as
\begin{align}
  & \dot p_k = e_k^\dagger X e_k ,  &X &= F\Big(\sum_{k=1}^N p_k e_k e_k^\dagger \Big)\\
  &\dot e_k = B e_k,  &B &= \sum_{k\neq l} \frac{e_k^\dagger X e_l}{p_k-p_l}e_ke_l^\dagger .
\end{align}
The matrices $E_{kk}=\mathrm{i}e_ke_k^\dagger \in \SU(N)$ forming the eigenbasis of $P$ are quantized analogues of the level curves of the stream function $\psi$.

We now apply the Lie theory machinery to obtain the dynamics of $W_s$ and $W_r$.
From the definition of $W_s$ we have
\begin{equation}
\begin{aligned}
\dot{W_s} &= \dfrac{d}{dt}\left(E\mathrm{diag}(E^\dagger W E)E^\dagger\right)\\
&=[\dot{E}E^\dagger,W_s] - \Pi_P([\dot{E}E^\dagger,W_r]),
\end{aligned}
\end{equation}
where we used $\Pi_P(\dot{W})=0$ and $\dot{E}E^\dagger=-E\dot{E^\dagger}$. 
The dynamics for $P$ is similar
\[
\dot{P} = [\dot{E}E^\dagger,P] + E\dot{\Lambda} E^\dagger.\]
Hence, a formula for $\dot{E}E^\dagger$ is needed.
But we know that the dynamics of $P$ can be orthogonally decomposed as
\[
\dot{P} = \Pi_P^\perp(\Delta_N^{-1}[P,\Delta P]) + \Pi_P(\Delta_N^{-1}[P,\Delta P]),\]
so 
\[[\dot{E}E^\dagger,P]= \Pi_P^\perp(\Delta_N^{-1}[P,\Delta P]).\]
Notice that $\dot{E}E^\dagger$ can be taken in $\Stab_P^\perp$.
In fact, the dynamics of $W_s$ remains the same for any $\dot{E}E^\dagger+S$, where $S\in\Stab_P$.
The map
\[[\cdot,P]:\Stab_P^\perp\rightarrow\Stab_P^\perp\]
is invertible so $\dot{E}E^\dagger$ is uniquely determined in $\Stab_P^\perp$.
In conclusion we have derived the following result for the dynamics of $W_s$ and $W_r$.
\begin{theorem}\label{thm:Ws_Wr_daynamics}
Let $W = W(t)$ and $P=P(t)$ be the vorticity and stream matrix for a solution to the Euler-Zeitlin equations \eqref{eq:Euler_eqn_qant1}.
Let $W_s$ and $W_r$ respectively be the orthogonal projections of $W$ onto $\Stab_P$ and its orthogonal complement as in~\eqref{eq:canonical_splitting}.
Then $W_s$ and $W_r$ satisfy the following system of equations
\begin{equation}\label{eq:Ws_Wr_equations}
\begin{aligned}
\dot{W_s} &= [B,W_s] - \Pi_P[B,W_r]\\
\dot{W_r} &= -[B,W_s] +  \Pi_P[B,W_r] + [P,W_r],
\end{aligned}
\end{equation}
where $P=\Delta_N^{-1}(W_s+W_r)$ and $B$ is the unique solution in $\Stab_P^\perp$ to
\begin{equation}\label{eq:B_def}
  \left[B,P\right] = \Pi_P^\bot\Delta_N^{-1}[P,W_r].
\end{equation}
\end{theorem}
From \autoref{thm:Ws_Wr_daynamics} we can deduce properties of $W_s$ and $W_r$.
First, if $W_r=0$ then $B\in\Stab_P\cap\Stab_P^\perp$ so $B=0$. 
Conversely, if $B=0$ then $\dot{W_s}=0$ and 
\begin{equation}\label{eq:euler_topography}
  \dot{W_r}=[\Delta_N^{-1}(W_s+W_r),W_r]  
\end{equation}
Hence, in that case $W_s$ plays the role of a fixed topography, and $W_r$ satisfies the Euler-Zeitlin equation with forcing~\eqref{eq:euler_topography}.
From equation \eqref{eq:B_def} we deduce that $B=0$ also implies
\begin{equation}\label{eq:cond_WsWr}
  \Tr(\Delta_N^{-1}W_s[\Delta_N^{-1}W_r,W_r])=0.
\end{equation}

Another observation is that if $[B,W_s]=0$ then $B=0$ so $\dot{W_s}=0$, and vice versa if $\Pi_P[B,W_r]=0$ then again equation \eqref{eq:cond_WsWr} holds.
This means that it is possible to have an evolution of the eigenvectors of $P$ without any change of the eigenvalues, but not the other way around.

\begin{remark}
  The projection $\Pi_P:\SU(N)\rightarrow\Stab_P$ has rank $N$ in the generic case.
  Hence, the dynamics of $W_s$ in the moving frame $E_{kk}$ can be described by only $N$ components, i.e., its eigenvalues. 
  Therefore, the vorticity splitting can be interpreted as reduced dynamics.
  As we shall see in \autoref{sec:vort_cont} below, the projection $\Pi_P$ is a quantized version of a mixing operator.
  Such operators was used by \citet{Shn1993} to characterise stationary flows.
\end{remark}

\subsection{Energy and enstrophy splitting}
Let us now study how energy and enstrophy relate to the canonical splitting~\eqref{eq:canonical_splitting}.
Since $\Tr(PW_r)=0$, the energy, corresponding to the \emph{energy norm}, fulfils
\begin{align*}
H(W) &= \frac{1}{2}\Tr\big(W_s\Delta_N^{-1}(W_s+W_r)\big)\\
&= \frac{1}{2}\Tr(W_s\Delta_N^{-1}W_s)-\frac{1}{2}\Tr(W_r\Delta_N^{-1}W_r).
\end{align*}
Yet, the enstrophy, corresponding to the \emph{enstrophy norm}, fulfils
\[E(W) = -\Tr(W_s^2) - \Tr(W_r^2).\]
This gives the interesting relations
\begin{equation}\label{eq:energy_enstrophy_splitting}
\begin{aligned}
H(W) &= H(W_s) - H(W_r)\\
E(W) &= E(W_s) + E(W_r).
\end{aligned}
\end{equation}
Notice that $H(W_s)$ is always larger than $H(W)$ and that the energy of $W_s$ and $W_r$ have to increase or decrease at the same rate. 
On the other hand, if the enstrophy of $W_s$ decrease with a rate then the enstrophy of $W_r$ must increase with the same rate.
The canonical splitting thus coheres with Kraichnan's~(\citeyear{Kr1967}) description of an inverse energy cascade and a forward enstrophy cascade.


The energy-enstrophy splitting \eqref{eq:energy_enstrophy_splitting} has a geometric interpretation.
It says that $W$ and $W_r$ are orthogonal in the energy norm, but $W_s$ and $W_r$ are orthogonal in the enstrophy norm.
Consider the plane spanned by $W$ and $W_r$, and let $H_r=H(W_r)$ and $H_0=H(W)$.
Then, since $W_r = (0,\sqrt{H_r})$ and $W=(\sqrt{H_0},0)$ in this plane, energy correspond to the Euclidean norm on~$\Rr^2$.
We want to express the enstrophy norm relative to the energy norm.
First observe that $W_s = (\sqrt{H_0},-\sqrt{H_r})$. 
The positive definite matrix $G$ for the enstrophy inner product restricted to the $(W,W_r)$-plane can be written $G=C^\top C$, where the matrix $C\in\Rr^{2\times 2}$ is determined by $C\cdot (0,\sqrt{H_r}) = (0,\sqrt{E_0}\sin\alpha)$ and $C\cdot (\sqrt{H_0},0) = \sqrt{E_0}(\cos\alpha,\sin\alpha)$.
Here, $\alpha$ is the angle between $W$ and $W_s$ in the enstrophy norm, and $E_0=E(W)$.
Then we have
\begin{equation}
G = \begin{bmatrix}
\dfrac{E_0}{H_0} & \dfrac{E_0}{\sqrt{H_r}\sqrt{H_0}}\sin^2\alpha \\
\dfrac{E_0}{\sqrt{H_r}\sqrt{H_0}}\sin^2\alpha  & \dfrac{E_0}{H_r}\sin^2\alpha 
\end{bmatrix}.
\end{equation}

\begin{proposition}
Let $W\neq 0$. Then
\begin{equation}\label{eq:en_ens_ineq1}
0< H(W)\leq H(W_s) < E(W_s)\leq E(W).
\end{equation}
Moreover, with notation as above
\begin{equation}\label{eq:en_ens_ineq2}
\begin{aligned}
& H_r\leq E_r=E_0 \sin^2\alpha \\
&N^2 H_r \geq E_r = E_0 \sin^2\alpha.
\end{aligned}
\end{equation}
Hence, if $\sin\alpha\neq 0$ then
\begin{equation}\label{eq:ratio_sin_r}
\dfrac{E_0}{N^2}\leq \dfrac{H_r}{\sin^2\alpha}\leq E_0.
\end{equation}
\end{proposition}

\proof
First notice that $W_s=0$ if and only if $W=0$ since $\sqrt{H(\cdot)}$ is a norm.
The inequalities \eqref{eq:en_ens_ineq1} then follow from \eqref{eq:energy_enstrophy_splitting} and the fact that the enstrophy is always larger than the energy.
To get the second inequality of \eqref{eq:en_ens_ineq2}, we use that the largest eigenvalue of the discrete Laplacian $\Delta_N$ is $(N-1)N$.
\endproof

\begin{remark}
In the limit $N\rightarrow\infty$ the ratio $\sin^2\alpha/H_r$ in \eqref{eq:ratio_sin_r} is potentially unbounded. 
It could happen, and in fact does happen, that the enstrophy norm of $W_r$ is far from zero, while its energy norm tends to zero.
This corresponds to $W_r$ being shifted towards small scales while not decreasing its enstrophy.
It is a manifestation of Kraichnan's theory of forward enstrophy and inverse energy cascades.
\end{remark}

\begin{figure*}
  \hspace{-3.0ex}
  \includegraphics[width=1.065\textwidth]{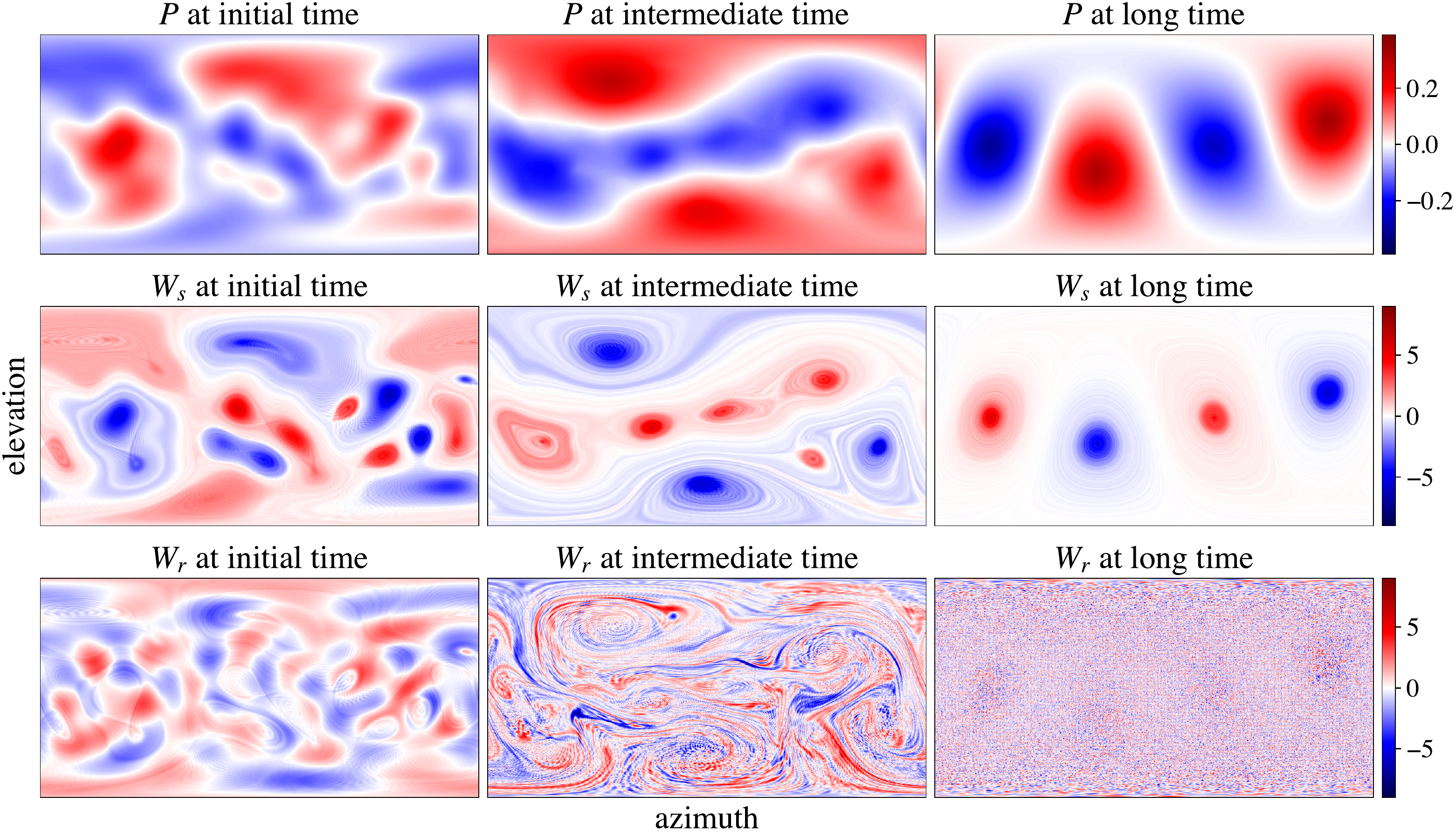} 
  \caption{Vanishing momentum simulation.
  Progression of the stream matrix $P$ and the components $W_s$ and $W_r$ for the same simulation as in \autoref{fig:evolution1}.
Initially $W_s$ and $W_r$ are similar in nature, but they evolve so $W_s$ traps the large-scale condensates whereas $W_r$ captures the small-scale fluctuations.
}\label{fig:Vorticity}
\end{figure*}

\section{Dynamically emerging scale separation}\label{sec:vort_numerics}
In this section we give numerical evidence that the canonical vorticity splitting~\eqref{eq:canonical_splitting} capture the dynamics of scale separation.
We also provide some theoretical arguments in support.
To give a complete mathematical proof that fully explains the numerical observations is a challenge not addressed in this paper.


A principal motivation for studying the vorticity splitting $W = W_s + W_r$ is that unsteadiness precisely means non-vanishing of $W_r$.
From an analytical point of view, $W_s$ represents a projection of $W$ onto a smoother subspace.
Indeed, the relation via the Laplace operator between $P$ and $W$ says that $P$ admits two more spatial derivatives than $W$.
Hence, since $W_s$ is related to $P$ via a polynomial relationship, $W_s$ is in general more regular than $W$.
Vice versa, $W_r$ contains the rougher part of $W$.
The tempting conjecture is that $W_s$ represents the low-dimensional, large-scale dynamics, whereas $W_r$ represents the noisy, small-scale dynamics. 
To assess this conjecture, we present two numerical simulation, both with randomly generated, smooth initial data.
These two simulations represent the two generic behaviours described by \citet{MoVi2020}: formation of either 3 or 4 coherent vortex blob formations, strongly correlated to the momentum-enstrophy ratio.\footnote{We have run many more simulation with randomly generated initial conditions. 
The two simulations presented here capture the universal behaviour in all simulations.}\

\begin{figure}
\hspace{-3.0ex}
\includegraphics[width=1.025\linewidth]{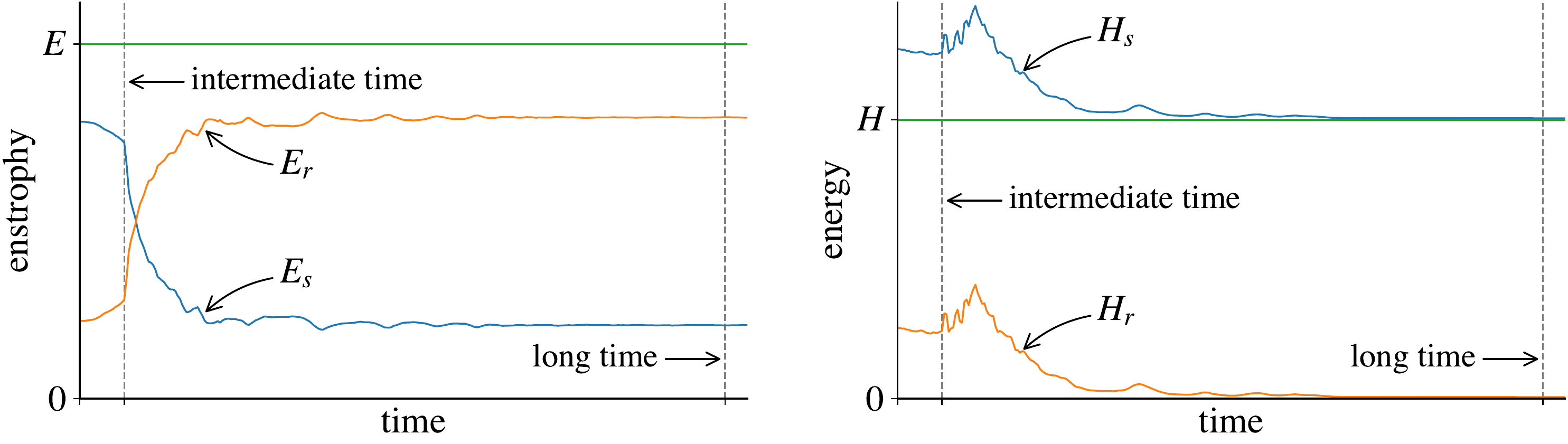}
\caption{
Vanishing momentum simulation.
Evolution of the decomposed enstrophies $E_s$ and $E_r$ (left), and decomposed energies $H_s$ and $H_r$ (right).
The dashed, vertical lines indicate the sample times in Figures~\ref{fig:evolution1},~\ref{fig:Vorticity},~\ref{fig:Ene_spec}.
The energy $H_r$ decays almost to zero, so that most of the energy is contained in $H_s$ (reflecting the inverse energy cascade).
On the other hand, the enstrophy $E_r$ increases over time (reflecting the forward enstrophy cascade). 
}\label{fig:E_H_evolution} 
\end{figure}

\begin{figure}
\hspace{-3.0ex}
\includegraphics[width=1.025\linewidth]{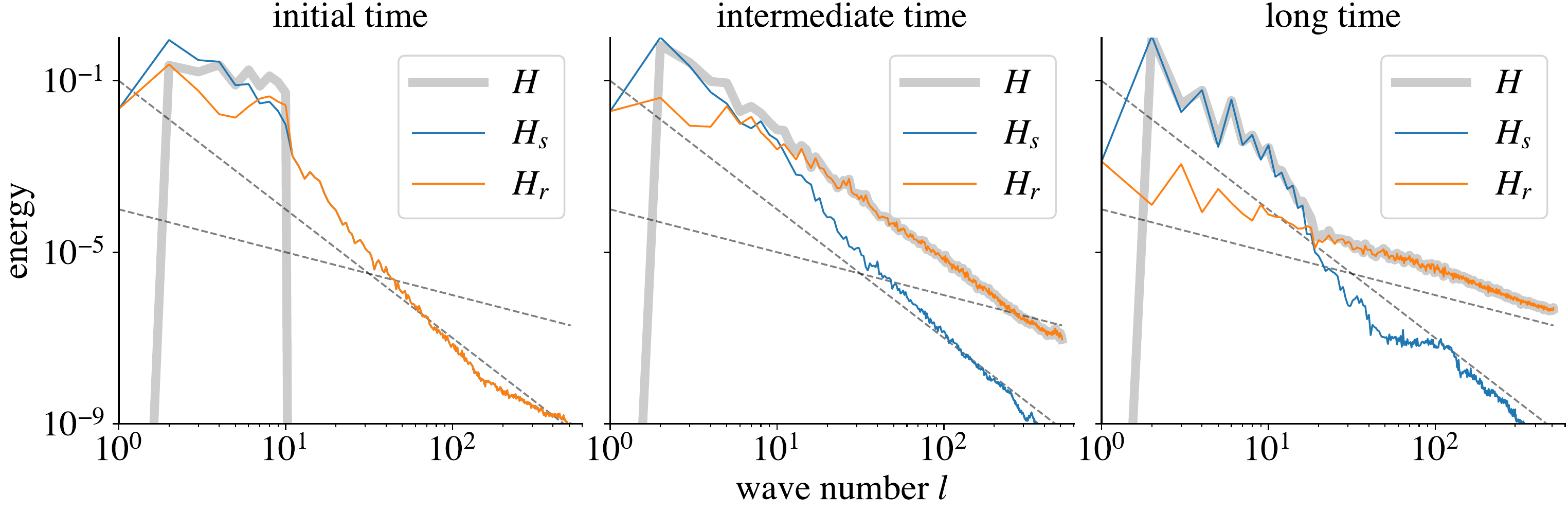}
\caption{Vanishing momentum simulation. 
Spectrum in log-log scale for the energies $H$, $H_s$, and $H_r$ at the initial (left), intermediate (middle), and long time (right).
The dashed lines indicate the slopes $l^{-3}$ and $l^{-1}$.
The slope of $H_s$ is almost settled at the intermediate time.
The slope of $H_r$ take much longer to settle.
At long time, the broken line spectrum of $H$ is captured well by the components $H_s$ and $H_r$, which themselves have almost the same average slope at each scale.
}\label{fig:Ene_spec}
\end{figure}

\subsection{Vanishing momentum simulation}
This simulation starts from smooth, randomly generated initial data.
Each spherical harmonic coefficient $\omega^{lm}$ in the range $2\leq l \leq 10$ was drawn from the standard normal distribution.
Remaining coefficients were set to zero.
The data was then transformed to a vorticity matrix of size $N=512$ (see \citet{MoVi2020} for details and explicit formulae).
The so generated initial conditions are available in the supplementary material for reproducibility.


For time discretization of the Euler-Zeitlin equations~\eqref{eq:Euler_eqn_qant1} we use the second order \emph{isospectral midpoint method} \citep{ModViv2019a}.
The time step length is $1.2239\cdot 10^{-4}\,$s.
This corresponds to $0.2$ time units in Zeitlin's model which scales time by $4\sqrt{\pi}/N^{3/2}$.

To visualise the fluid motion stages, we sample at initial time ($t=0$), at intermediate time during mixing ($t=13\,$s), and at long time well after the large scale condensates are formed ($t=318\,$s).
The vorticity matrices at these outputs are then transformed to functions in azimuth-elevation coordinates.
The result is shown in~\autoref{fig:evolution1}.

Due to vanishing momentum ($\omega^{1m}=0$ in the initial data), integrability theory for Hamiltonian dynamics on the sphere suggests that 4 vortex blobs should appear \citep{MoVi2020,MoVi2020b}.
Indeed they do appear, and then pass into quasi-periodic orbits.
The first movie of the supplementary material captures the entire process.
On top of the large scale condensates, a noisy, fine scale structure emerge.
In essence we see a separation of scales.

\autoref{fig:Vorticity} displays azimuth-elevation fields for the stream matrix $P$ and the vorticity matrix components $W_s$ and $W_r$ at the same sampled output times.
After some time of initial mixing the large scales of the vorticity are all contained in $W_s$, whereas $W_r$ collects the small-scale fluctuations.
The long time $W_r$ state resembles noise, but not completely uniform.
In fact, the non-uniform character captures the quasi-periodic dynamics since $\dot W = [P,W_r]$.
An open problem is to model the noise by a stochastic process.

\autoref{fig:E_H_evolution} shows the evolution of the energy and enstrophy of $W_s$ and $W_r$. 
Over time $E_r$ increases whereas $H_r$ decreases.
Also notice that $E_s$, $E_r$, $H_s$, and $H_r$ fulfil the energy and enstrophy relations~\eqref{eq:energy_enstrophy_splitting}.
The residual vorticity $W_r$ decreases in energy norm but increases in enstrophy norm.
This is a quantification of Kraichnan's~(\citeyear{Kr1967}) qualitative description.

The scale separation of the vorticity is even more clear in the spectral domain.
\autoref{fig:Ene_spec} contains energy spectra for $W$, $W_s$, and $W_r$ at the sampled output times.
The energy level $H(l)$, corresponding to the wave-number $l=1,\ldots, N$, contains the energy of the modes for the spherical harmonics $Y_{lm}$ with $m=-l,\ldots,l$.
We notice that the energy spectrum of $W$ is similar in nature to that described by \citet{BoEc2012}.
Indeed, the ``broken line'' slope in the energy spectrum of $W$ originate from an $l^{-3}$ slope of $W_s$ and an $l^{-1}$ slope of $W_r$.
Thus, the vorticity splitting yields a scale separation of the vorticity field that exactly reflect the broken line spectra previously observed in numerical simulations and empirical observations.
The broken line spectrum is not reached at the intermediate time, before mixing has settled.


\subsection{Non-vanishing momentum simulation}

\begin{figure*}
  \hspace{-3.0ex}
  \includegraphics[width=1.065\textwidth]{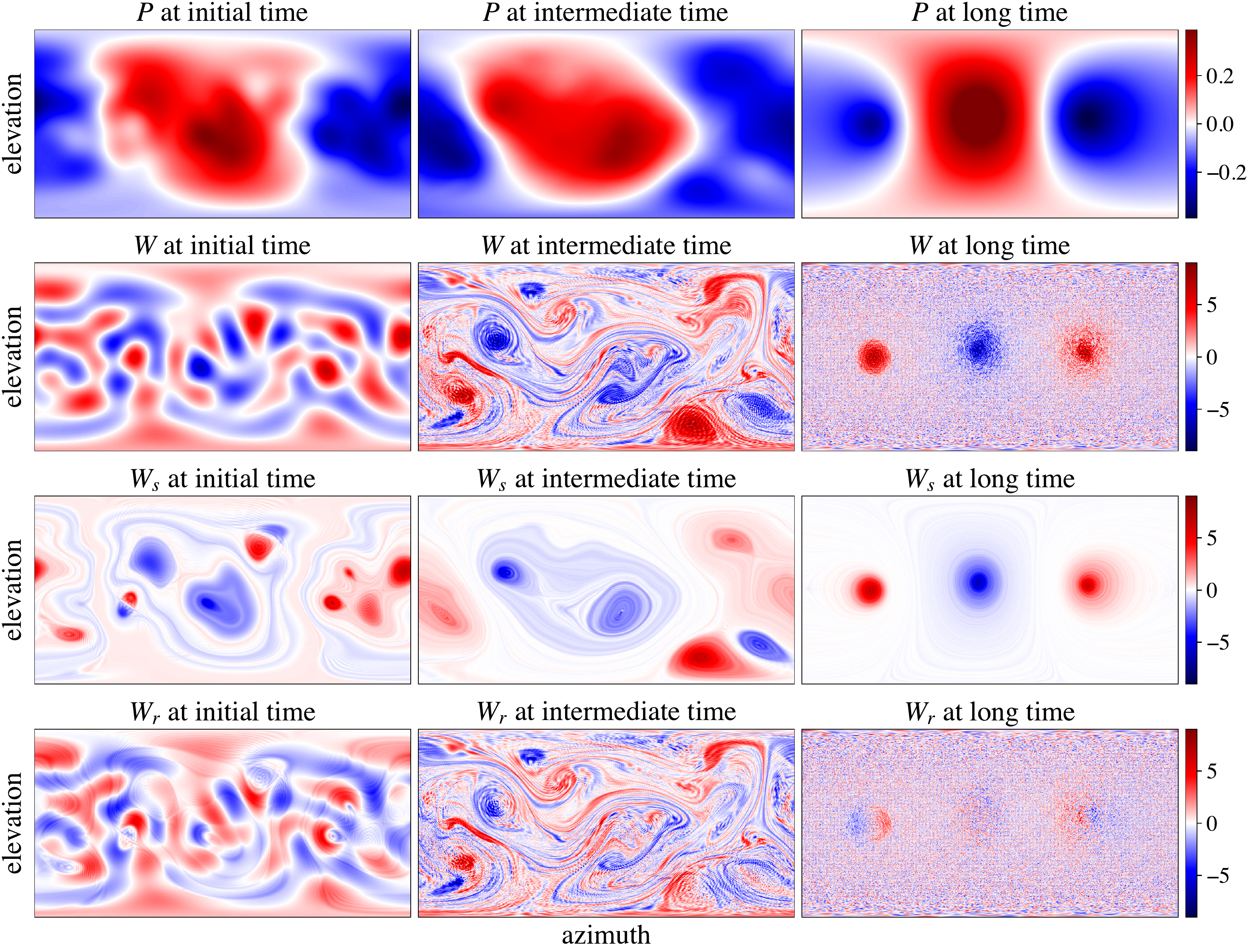} 
  \caption{Non-vanishing momentum simulation. Progression of the stream matrix $P$ and the components $W_s$ and $W_r$ of the vorticity matrix $W$.
Initially $W_s$ and $W_r$ are similar in nature, but they evolve so that $W_s$ contains the large-scale condensates whereas $W_r$ contains the small-scale fluctuations.
}\label{fig:Vorticity2}
\end{figure*}

In this simulation the initial data was generated much like before, but now the range of non-zero spherical harmonics coefficients $\omega^{}lm$ is $1 \leq l \leq 10$.
This means total angular momentum is no longer zero.
For reproducibility also of this simulation, the generated initial conditions are available in the supplementary material.
Time discretization, step size lengths, etcetera are selected as in the previous simulation.


\autoref{fig:Vorticity2} shows azimuth-elevation fields corresponding to the stream matrix $P$, the vorticity matrix $W$, and the components $W_s$ and $W_r$, sampled at initial time ($t=0$), intermediate time ($t=13$s), and long time ($t=344$s).
The entire motion is captured in the second movie of the supplementary material.
Three vortex blobs emerge. 
The formation of these, from initial data with non-vanishing momentum, is again predicted and demonstrated by \citet{MoVi2020}.
It reflects integrability of low-dimensional Hamiltonian dynamics on the sphere.
As before, the large scale vorticity patterns are contained in the $W_s$ component. 
The $W_r$ component swiftly develop noisy fluctuations.
At long time it is less uniform than in the vanishing momentum simulation.
The reason is that the 3 blobs here move faster than the 4 blobs in the previous simulation.

\begin{figure}
\hspace{-3.0ex}
\includegraphics[width=1.025\linewidth]{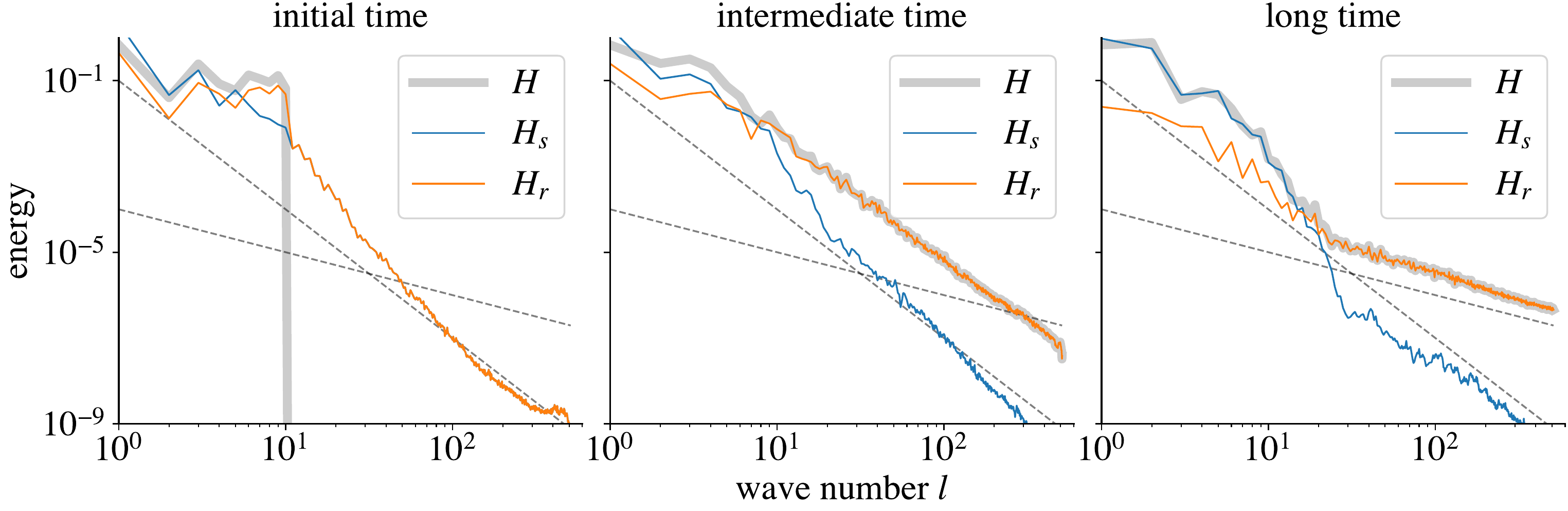}
\caption{Non-vanishing momentum simulation. 
Spectrum in log-log scale for the energies $H$, $H_s$, and $H_r$ at the initial (left), intermediate (middle), and long time (right).
The dashed lines indicate the slopes $l^{-3}$ and $l^{-1}$.
The slope of $H_s$ is almost settled at the intermediate time.
The slope of $H_r$ take much longer to settle.
At long time, the broken line spectrum of $H$ is captured well by the components $H_s$ and $H_r$, which themselves have almost the same average slope at each scale.
}\label{fig:Ene_spec2}
\end{figure}

The scale separation of the vorticity is again evident from the energy spectrum of~$W$.
\autoref{fig:Ene_spec2} gives energy spectra for the three vorticity fields $W$, $W_s$, and $W_r$.
The results are analogous to those in \autoref{fig:Ene_spec}.


\subsection{Stream function versus vorticity branching and blobs}

In the literature on 2D turbulence steady solutions are often characterised by a functional relation between stream function and vorticity.
Branching in such relations has been used as a measure of unsteadiness~\citep{dri}.

\begin{figure}
\hspace{-3.0ex}
\includegraphics[width=1.025\linewidth]{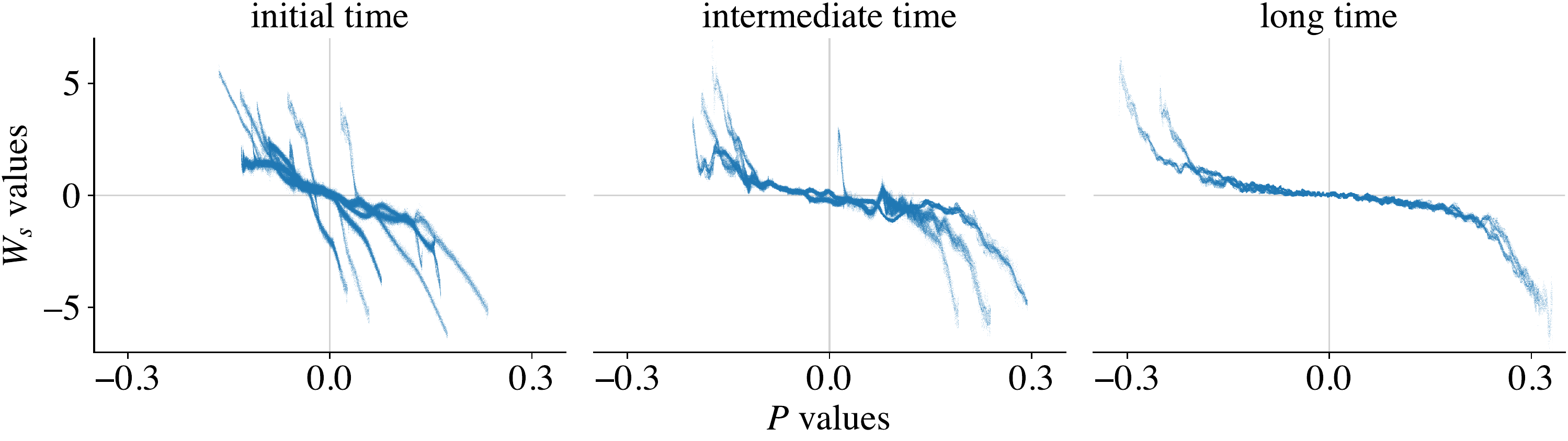}
\caption{Vanishing momentum simulation. 
Values of $P$ versus values of $W_s$.
The end of each branch correspond to a blob in the $W_s$ plot in \autoref{fig:Vorticity}.
Upward if the blob is positive, otherwise downward.
For example, at intermediate time, the sharp upward branch close to the $y$-axis matches the small positive blob of $W_s$ in the lower right corner of the corresponding plot in \autoref{fig:Vorticity}.
}\label{fig:branching4blobs}
\vspace{2ex}
\hspace{-3.0ex}
\includegraphics[width=1.025\linewidth]{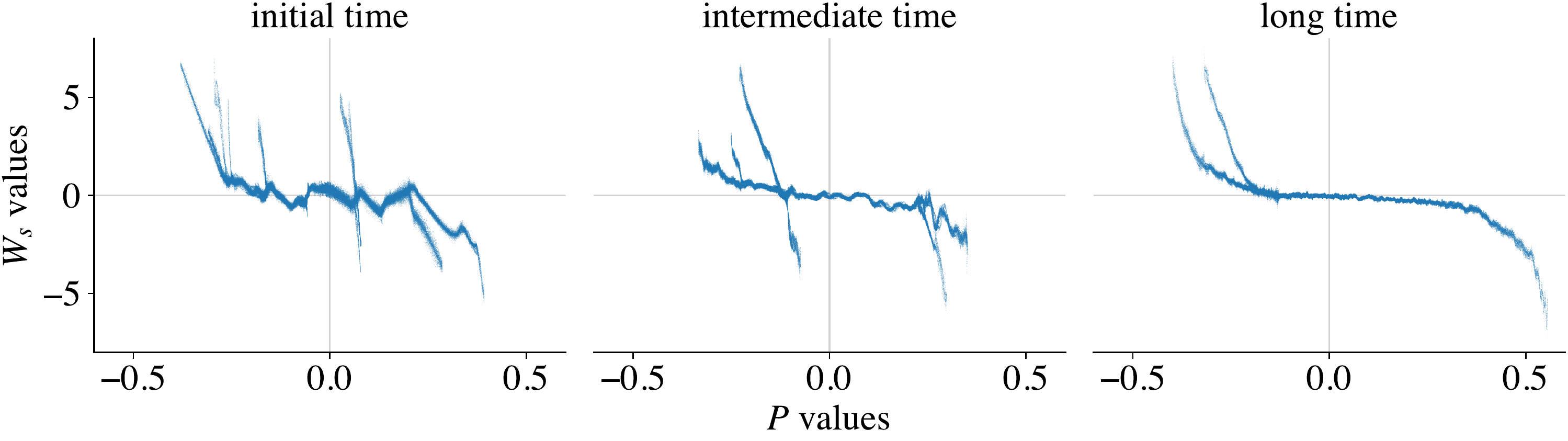}
\caption{Non-vanishing momentum simulation.
Values of $P$ versus values of $W_s$. 
The end of each branch correspond to a blob in \autoref{fig:Vorticity2}.
For example, at initial time, a careful study reveals that the steep, overlapping branches just to the right of the $y$-axis match the two positive-negative blob-pairs seen in the corresponding plot in \autoref{fig:Vorticity2}.
}\label{fig:branching3blobs}
\end{figure}

Since $W = W_s$ if and only if $W$ is a steady solution, it is natural to consider scatter plots between values of $P$ and $W_s$.
Such diagrams are given at the initial, intermediate, and long times, in \autoref{fig:branching4blobs} for the vanishing momentum simulation, and in \autoref{fig:branching3blobs} for the non-vanishing momentum simulation.

The following interpretation of branches is primal over interpretations related to unsteadiness.
Each branch represents and characterises a specific blob in $W_s$.
Upward branches represent blobs with positive values.
Vice versa for downward branches.
It is particularly clear at the long times, where there are fewer blobs.
But the interpretation is valid also at the initial and intermediated times.
This is revealed by carefully comparing the branching diagrams with the values of $P$ and $W_s$ in \autoref{fig:Vorticity} and \autoref{fig:Vorticity2}.
The end of each specific branch (sometimes they overlap) is then readily identified with a specific blob.
How steep the branch is corresponds to $d W_s/dP$.
This can be used to determine the shape of the blob, assuming axi-symmetry.
For example, in \autoref{fig:branching3blobs} at long time, the steeper of the two upward branches correspond to the left-most, sharper of the two positive blob in \autoref{fig:Vorticity2}.
The two negative blobs at long time in \autoref{fig:Vorticity} are almost indistinguishable, which is reflected as overlapping downward branches in \autoref{fig:branching4blobs}.

\section{Canonical splitting of the vorticity function}\label{sec:vort_cont}
In this section we map our results for the Euler-Zeitlin equations to the original Euler equations~\eqref{eq:euler_eqn_1}.
Indeed, all the concepts needed in the canonical splitting for the Euler-Zeitlin equations have classical counter-parts; some are listed in \autoref{tbl:dictionary}.
However, one has to be careful to rigorously define these concepts: the transition from the quantized to the classical equations is valid only in a weak sense.
Mathematically, the correct framework is $L^\infty$ weak-star convergence.
Formally, we may nevertheless proceed as follows, keeping in mind that concepts are transferable only in the weak sense.
The key ingredient is that the projection $\Pi_P$ onto the stabilizer of $P$ corresponds to averaging along the level-sets of the stream function $\psi$.
This gives the canonical splitting for the vorticity function via the projector $\Pi_\psi$ as
\begin{equation}\label{eq:splitting_continuous}
\omega = \Pi_\psi(\omega) + \Pi_\psi^\bot(\omega) = \omega_s + \omega_r .
\end{equation}
The projector $\Pi_\psi$ is time dependent, since the level curves of $\psi$ change with time.

Let us now proceed with more mathematical details on this constructions.
First, recall again Euler's equations in vorticity formulation
\begin{equation}\label{eq:euler_eqn_cont}
\dot{\omega} =\lbrace\psi,\omega\rbrace, \qquad
\Delta \psi = \omega.
\end{equation}
To define the continuous analogue of the vorticity matrix splitting, we have to understand the equations~\eqref{eq:euler_eqn_cont} in the weak sense.
Indeed, in general the projections $\Pi_\psi$ and $\Pi_\psi^\bot$ cannot preserve the smoothness of $\omega$.
But for any $p\in[1,\infty]$ they are continuous operators from $C^0(\Ss^2)$ to $L^p(\Ss^2)$ with operator norm one.
Since continuous functions are dense in $L^p(\Ss^2)$, we can extend $\Pi_\psi$ and $\Pi_\psi^\bot$ to continuous operators on $L^\infty(\Ss^2)$.
This result fits well with the global well-posedness of equations \eqref{eq:euler_eqn_cont}, which precisely requires a vorticity function in $L^\infty$ \citep{Yu1963,MaPu2012}.

To show this, let us first give equations \eqref{eq:euler_eqn_cont} in the weak sense.
For any $p \geq 2$, if $\omega\in L^p(\Ss^2)$ then $\psi\in W^{2,p}(\Ss^2)\subset H^1(\Ss^2)$.
We define the weak Poisson bracket as
\[
\int_{\Ss^2}\lbrace\psi,\omega\rbrace\phi =-\int_{\Ss^2}\omega\lbrace\psi,\phi\rbrace ,
\]
for any test function $\phi\in C^\infty(\Ss^2)$.
Hence, we define the stabilizer of $\psi$ as
\[\Stab_\psi:=\lbrace f\in L^2(\Ss^2) \mid \lbrace f,\psi\rbrace=0\rbrace.\]
Next we define the $L^2$ orthogonal projection $\Pi_\psi$ onto $\Stab_\psi$.
\begin{proposition}
If $p\geq 2$ and $\psi\in W^{2,p}(\Ss^2)$
then $\Stab_\psi$ is a closed subspace of $L^2(\Ss^2)$.
\end{proposition}
\proof
Let $\lbrace f_n\rbrace\subset \Stab_\psi$, such that $f_n\rightarrow f$ in $L^2$, for $n\rightarrow\infty$.
We want to show that $f\in\Stab_\psi$.
Let $\phi$ be a function in $C^1(\Ss^2)$, then we get
\[
|\int \lbrace f,\psi\rbrace\phi |= |\int \lbrace \phi,\psi\rbrace f |=|\int \lbrace \phi,\psi\rbrace (f-f_n) |\leq \|\lbrace \phi,\psi\rbrace\|_0\|f-f_n\|_0\rightarrow 0, \]
for $n\rightarrow\infty$.
\endproof

\begin{table}
\centering
\begin{tabular}{rll} 
  & \textbf{Euler's equations} & \textbf{Zeitlin's model} \\ 
\emph{Lie group} & symplectic diffeomorphisms & $\mathrm{SU}(N)$ \\
\emph{Lie algebra} & divergence free vector fields & $\SU(N)$ \\
\emph{Phase space} & $L^\infty$ functions & $\SU(N)^*\simeq \SU(N)$ \\
\emph{Strong norm} & supremum norm $\norm{\cdot}_{L^\infty}$ & spectral norm \\
\emph{Energy norm} & Sobolev norm $\norm{\cdot}_{H^{-1}}$ & $\sqrt{-\operatorname{Tr}(PW)}$ \\
\emph{Enstrophy norm} & $L^2$ norm $\norm{\cdot}_{L^2}$ & Frobenius norm \\
\emph{Canonical splitting} & averaging on level-sets of $\psi$ & projection onto stabilizer of $P$ \\
\emph{$z$-axis symmetry} & $\omega$ is zonal & $W$ is diagonal \\
\end{tabular}
\caption{Dictionary between Euler's and Zeitlin's models of hydrodynamics.}\label{tbl:dictionary}
\end{table}

As anticipated above, the operator $\Pi_\psi$ has an explicit form when evaluated on continuous functions.
To state it, we first make the following assumption on the critical points of $\psi$.
\begin{ass}\label{ass:generic_psi}
Let $\psi\in C^1(\Ss^2)$ be the stream function. 
Then the critical points of $\psi$ define a set of zero Lebesgue measure on $\Ss^2$, such that it is never dense in any neighbourhood of one of its points.
\end{ass}
We say that $\psi$ is \textit{generic} whenever it satisfies Assumption~\ref{ass:generic_psi}.
Consider now some $f\in C^1(\Ss^2)$.
Then $f\in\Stab_\psi$ if and only $\nabla f$ and $\nabla\psi$ are parallel.
Since we take $\psi$ to be generic, the points where $\{ x\mid \nabla\psi(x) = 0\}$ lie on a set of zero measure, nowhere dense.
Therefore, since $f$ is continuous, $f\in\Stab_\psi$ if it is constant on the connected components of the level curves of $\psi$.
Then the projection of $f$ onto $\Stab_\psi$ can be defined by evaluating $f$ on the level curves of $\psi$, i.e., the streamlines.
Let $\gamma$ be a connected component of a streamline, then define the projection $\Pi_\psi:C^1(\Ss^2)\rightarrow\Stab_\psi$ as
\begin{equation}\label{ee:Pi_s_defn}
\Pi_\psi(f)\big|_{\gamma}=\frac{1}{\mbox{length}(\gamma)}\int_\gamma f ds.
\end{equation}
In the limit when $\gamma$ approaches a single point, clearly $\Pi_\psi(\omega)_{|\gamma}=f(\gamma)$. 

The operator $\Pi_\psi$ does not, in general, preserve continuity of $f$.
Indeed, consider a \textit{bifurcation saddle point} $x\in\Ss^2$, i.e.\ $x$ is a saddle point of $\psi$ such that the streamline passing through $x$ contains a bifurcation point.
We then have the following result.
\begin{proposition}
Let $\psi$ be generic and $\Pi_\psi$ the projector as defined in \eqref{ee:Pi_s_defn}. 
Then, if $x\in\Ss^2$ is a bifurcation saddle point for $\psi$, there exists $f\in C^1(\Ss^2)$, such that $\Pi_\psi(f)$ is discontinuous at the streamline passing through $x$.
Vice versa, given $f\in C^1(\Ss^2)$, if $\Pi_\psi(f)$ is discontinuous at some point $x\in\Ss^2$, then the streamline passing through $x$ contains a bifurcation saddle point.
\end{proposition}
\proof
(Sketch) The issue about the continuity of $f$ can be treated locally.
Hence, let us work in Cartesian coordinates.
Let $x\in\Ss^2$ be a bifurcation saddle point for $\psi$ and $\gamma$ the streamline passing through $x$.
Then, let $\beta$ be a curve intersecting $\gamma$ only in $x$ and let $f$ be a smooth function positive at one side of $\beta$ and negative at the other one, such that $\int_\gamma f ds=0$.
Then being $x$ a bifurcation point, for any neighbourhood $U$ of $x$, there exist streamlines totally contained in one or another side of $\beta$.
Then, the average of $f$ on those streamlines is either strictly positive or negative, creating a discontinuity at $\gamma$. 

Vice versa, let $f\in C^1(\Ss^2)$, such that $\Pi_\psi(f)$ is discontinuous at some point $x\in\Ss^2$.
Then, the streamline passing through $x$ cannot be homeomorphic to any of those in some tubular neighbourhood.
Hence, the streamline passing through $x$ must contain a critical point for $\psi$, which also is a bifurcation saddle point.
\endproof
However, we have the following regularity for $\Pi_\psi$.
\begin{proposition}
For any $p\in[1,\infty]$, the operator $\Pi_\psi$ can be extended to a bounded operator with norm one on $L^p(\Ss^2)$.
\end{proposition}
\proof
Let us first notice that $\Pi_\psi$ is defined from $C^0(\Ss^2)$ to $L^1(\Ss^2)$, and satisfies $\|\Pi_\psi f\|_{L^1}\leq\|f\|_{L^1}$, for any $f\in C^0(\Ss^2)$.
Since $C^0(\Ss^2)$ is dense in $L^1(\Ss^2)$, it is possible to extend $\Pi_\psi$ to a bounded operator on $L^1(\Ss^2)$.
Secondly, $\Pi_\psi$ is well defined from the space of simple functions to $L^\infty(\Ss^2)$, and satisfies $\|\Pi_\psi f\|_{L^\infty}\leq\|f\|_{L^\infty}$, for any $f$ simple.
Since the space of simple functions is dense in $L^\infty(\Ss^2)$, it is possible to extend $\Pi_\psi$ to a bounded operator on $L^\infty(\Ss^2)$.
Furthermore, since $\Pi_\psi$ fixes the constant functions, it must be that $\|\Pi_\psi\|_{L^1}=\|\Pi_\psi\|_{L^\infty}=1$.
Hence, by the Riesz-Thorin theorem, we conclude that $\|\Pi_\psi\|_{L^p}=1$, for any $p\in[1,\infty]$.
\endproof

Hence, from now on, let us consider equations \eqref{eq:euler_eqn_cont} in the weak form, for $\omega\in L^\infty(\Ss^2)$.
It is clear that $\Pi_\psi^2=\Pi_\psi$.
Moreover, we can formally define the operator $\Pi_\psi$ via the kernel $$K(x,y) = \frac{1}{\mbox{length}(\gamma_x)}\delta_{\gamma_x}(y),\quad \text{for any}\; x,y\in\Ss^2$$ where $\gamma_x$ is the connected component of the streamline passing through $x$.
In this way we get that $\Pi_\psi$ is self-adjoint with respect to the $L^2$ inner product, i.e., for any $f,g\in C^\infty(\Ss^2)$
\begin{align*}
\int_{\Ss^2} f\Pi_\psi g  &=  \int_{\Ss^2} f(x)\int_{\Ss^2}K(x,y) g(y)(y)(x)\\
&= \int_{\Ss^2} g(y)\int_{\Ss^2}K(x,y) f(x)(x)(y)\\
&=\int_{\Ss^2} g\Pi_\psi f .
\end{align*}
By extension these equalities are valid for $f,g\in L^p(\Ss^2)$ whenever $p\in[1,\infty]$.

\begin{remark}
We notice that the operator $\Pi_\psi$, defined by the kernel $K(x,y)$, is a \textit{mixing operators} (or polymorphisms or bistochastic operators), as introduced by \citet{Shn1993,Shn2012}.
Such operators give rise to a partial ordering on $L^2(\Ss^2)$: for any $f,g\in L^2(\Ss^2)$, $f\dashv g$ if there exist a mixing operator, defined via the kernel $K$, such that $f=K\ast g$.
In his work, Shnirelman shows that stationary flows are characterised as \textit{minimal elements} of this ordering.
In a way, our work shows that it is enough to consider mixing operators of the form $\Pi_\psi$.
Within this class, $\omega$ is minimal if there exists a stream function such that $\Pi_\psi \omega=\omega$.
As we see next, this in turn implies that $\omega$ is stationary.
\end{remark}

\begin{proposition}
Let $\psi\in C^1(\Ss^2)$ be generic.
For $\omega\in L^\infty(\Ss^2)$ we then have
\begin{equation}
  \omega\in\Stab_\psi \iff \Pi_\psi \omega=\omega .
\end{equation}
\end{proposition}
\proof
We prove the result for $\omega\in C^1(\Ss^2)$, then conclude by extension.
Let $\omega\in\Stab_\psi$. 
Then, $\nabla \omega$ is parallel to $\nabla\psi$ almost everywhere.
Hence, the gradient of $\omega$ along any streamline must vanish, and so on each connected component it is constant.
By continuity of $\omega$ we deduce that $f$ must be constant also on the streamlines containing critical points.
Therefore, $\Pi_\psi\omega=\omega$.
Assume now that $\Pi_\psi\omega=\omega$. 
Then $\omega$ must be constant on each connected component of a streamline.
Hence, $\nabla \omega$ is orthogonal to the streamlines.
Since $\nabla\psi$ is also orthogonal to the streamlines, we conclude that $\lbrace \omega,\psi\rbrace=0$, i.e.\ $\omega\in\Stab_\psi$.
\endproof
 
We are now in position to derive continuous analogues of the results in \autoref{sec:vort_split} (which, remember, are based on Lie theory for matrices).
First, the stream function $\psi$ satisfies the equation
\begin{equation}\label{eq:euler_eqn_psi_cont}
\dot{\psi} =\Delta^{-1}\lbrace\psi,\Delta \psi\rbrace.
\end{equation}
This equation is not Hamiltonian. 
But we can split the right hand side into a Hamiltonian and non-Hamiltonian part via the projection $\Pi_\psi$
\begin{equation}\label{eq:euler_eqn_psi_cont2}
\dot{\psi} = \Pi_\psi^\bot\Delta^{-1}\lbrace\psi,\Delta \psi\rbrace + \Pi_\psi\Delta^{-1}\lbrace\psi,\Delta \psi\rbrace.
\end{equation}
Analogous to the quantized case, we seek a generator for $\Pi_\psi^\bot\Delta^{-1}\lbrace\psi,\Delta \psi\rbrace$.
That is, a function $b\colon \Ss^2\to\Rr$ such that
\begin{equation}\label{eq:b_def_eqn}
\lbrace b,\psi\rbrace = \Pi_\psi^\bot\Delta^{-1}\lbrace\psi,\Delta \psi\rbrace.
\end{equation}
It is clear that a necessary condition for the equation $\lbrace b,\psi\rbrace = f$ to have a solution $b$ is that $f\in \Stab_\psi^\perp$. 
Indeed, if $b\in C^1(\Ss^2)$ then $\lbrace b,\psi\rbrace\in \Stab_\psi^\perp$ since
\begin{equation}
\int_{\Ss^2} \lbrace b,\psi\rbrace g = -\int_{\Ss^2} \lbrace g,\psi\rbrace b = 0 \quad \text{for any} \; g\in\Stab_\psi .
\end{equation}
However, in general equation \eqref{eq:b_def_eqn} can be solved only where $\nabla\psi\neq 0$. 
Around the critical points of $\psi$ the gradient of $b$ is potentially unbounded.
Moreover, the right hand side in equation \eqref{eq:b_def_eqn} can be discontinuous at the level curves of $\psi$ containing saddle points of $\psi$.
Hence, $b$ can only be defined almost everywhere. 
Furthermore, we have the following:
\begin{proposition}
Let $f\in C^0(\Ss^2)$ and $\psi$ be generic. 
Then $f\in \Stab_\psi^\perp$ if and only if there exists $b$ almost everywhere smooth, such that $\lbrace b,\psi\rbrace = f$ on the set $\nabla\psi\neq 0$.
\end{proposition}
\proof The if part is clear. 
Let instead take $f\in \Stab_\psi^\perp$. 
Then, for any point $x\in\Ss^2$, we have to solve the PDE for $b$:
\begin{equation}\label{eq:PDE_of_B}
\nabla^\perp\psi\cdot\nabla b = f,
\end{equation}
where $\nabla^\perp\psi=x\times\nabla\psi$.
If $\nabla\psi$ does not vanish, equation \eqref{eq:PDE_of_B} can be solved by integration in the direction of $\nabla^\perp\psi$.
In the points where $\nabla\psi$ does not vanishes, $\nabla b$ is not defined by equation \eqref{eq:PDE_of_B}, and it can be unbounded around those points.
Hence, the field $b$ is almost everywhere smooth and satisfies $\lbrace b,\psi\rbrace = f$, where $\nabla\psi\neq 0$.
\endproof

%
%

\subsection{Dynamics of \texorpdfstring{$\omega_s$}{w_s} and \texorpdfstring{$\omega_r$}{w_r}}

To derive the dynamical equations for $\omega_s$, we cannot directly define the field $b$ corresponding to the quantized field $B$ above.
Instead, we consider the volume preserving vector field 
\[X[\psi]:=\Pi_\psi^\bot\Delta^{-1}\lbrace\psi,\Delta \psi\rbrace.\]
We note that $X$ corresponds to the infinitesimal action of a map $\varphi_t$ which transports $\psi$ by deforming its level curves. 
Hence, $\varphi_t$ acts naturally on $\Stab_\psi$.
Let us write $\Pi_\psi^t$ for the projection onto $\Stab_\psi$ at time $t$.
Then, for any point $x\in\Ss^2$, let 
\[
\gamma(t)=\lbrace y\mid \psi(y)=\psi(x)\mbox{ and $y$ in connected component of $x$}\rbrace ,
\]
and $d\hat{s_t}$ the normalized Lebesgue measure on $\gamma(t)$. 
We have then the formal identity
\begin{equation}
\begin{array}{ll}
\omega_s(t,x) &= \Pi_\psi^t\omega(t,x) \\
&= \int_{\gamma(t)}\omega(t,y) d\hat{s_t}(y) \\ 
&=  \int_{\gamma(0)}(\varphi_t^*\omega)(t,y) d\hat{s_0}(y)
\\
&= \Pi_\psi^0(\varphi_t^*\omega)(t,\varphi_t^{-1}(x)).
\end{array}
\end{equation}
Hence, for any test function $\phi\in C^\infty(\Ss^2)$
\begin{equation}
\begin{array}{ll}
&\dfrac{d}{dt}\int_{\Ss^2}\omega_s(t,x)\phi(x)  =
\dfrac{d}{dt}\int_{\Ss^2}\Pi_\psi^0(\varphi_t^*\omega)(t,\varphi_t^{-1}(x))\phi(x)  \\ 
&=\int_{\Ss^2}\left(\Pi_\psi^0(\varphi_t^*\mathcal{L}_X\omega)(t,\varphi_t^{-1}x) - \mathcal{L}_X\Pi_\psi^0(\varphi_t^*\omega)(t,\varphi_t^{-1}x)\right)\phi(x) \\
&=-\int_{\Ss^2}\omega(t,x) \mathcal{L}_X\Pi_\psi^t\phi(x) + \mathcal{L}_X\Pi_\psi^t\omega(t,x)\phi(x) \\
\end{array}
\end{equation}
where $\mathcal{L}_X$ is the Lie derivative, which simply acts on functions as $\mathcal{L}_Xf=X[f]$.
We notice from the previous calculations that $\mathcal{L}_X$ has to be evaluated only on elements in $\Stab_\psi$.
Hence, the time derivative of $\omega_s$ is well defined in the weak sense.

Let us now formally denote $X:=-\lbrace b,\cdot\rbrace$. 
Then, interpreting the Poisson bracket in the weak sense, we can write the dynamical system for $\omega_s$ and $\omega_r$ as
\begin{align*}
\dot{\omega_s} &= \lbrace b,\omega_s\rbrace - \Pi_\psi\lbrace b,\omega_r\rbrace\\
\dot{\omega_r} &= -\lbrace b,\omega_s\rbrace +  \Pi_\psi\lbrace b,\omega_r\rbrace + \lbrace\psi,\omega_r\rbrace\\
\lbrace b,\psi\rbrace &= \Pi_\psi^\bot\Delta^{-1}\lbrace\psi,\omega_r\rbrace,
\end{align*}
where $b$ is implicitly defined by the third equation and $\psi=\Delta^{-1}(\omega_s+\omega_r)$. 
We notice that the equations of motion for $\omega_s$ can also be written in a more compact form as
\[
\dot{\omega_s}=[\Pi_\psi,\mathcal{L}_X]\omega,
\]
where the square bracket is the commutator of operators.

Finally, notice that the energy and enstrophy splitting is valid also in the classical setting
\begin{align*}
H(\omega) &= H(\omega_s) - H(\omega_r)\\
E(\omega) &= E(\omega_s) + E(\omega_r).
\end{align*}

\section{Conclusions and outlook}\label{sec:outlook}

Based on the Euler-Zeitlin equations we have reported on a new technique for studying 2D turbulence via canonical splitting of vorticity.
In numerical simulations this splitting dynamically develops into a separation of scales.
These numerical results are supported by some theoretical evidence.
To develop a full mathematical understanding, even within the setting of Zeitlin's model, is an interesting open problem.
We have further presented mathematical foundations for a weak $L^\infty$ theory in the continuous setting, independent of Zeitlin's model.

As the numerical simulations so strikingly capture the scale separation process, and as the inverse relations of the corresponding energy-enstrophy splitting reflect the stationary theory of Kraichnan, it is likely that further numerical and theoretical studies of the canonical vorticity splitting shall unveil more details on the mechanism behind vortex condensation.
Furthermore, the splitting into large scales $\omega_s$ and small scales $\omega_r$ suggests to use these variables as a basis for \emph{stochastic model reduction} (\emph{cf.}~\citet{JaRiVa2015}) of the two-dimensional Euler equations, with $\omega_r$ modelled as multiplicative noise.

\textbf{Acknowledgements}{The computations were enabled by resources provided by the Swedish National Infrastructure for Computing (SNIC) at C3SE partially funded by the Swedish Research Council through grant agreement no. 2018-05973.}

\textbf{Funding}{
This work was supported by the Swedish Research Council (K.M., grant number 2017-05040); the Knut and Alice Wallenberg Foundation (K.M., grant number WAF2019.0201), (M.V., grant number KAW2020.0287); and the research grant Junior Visiting Fellowship by the Scuola Normale Superiore of Pisa, Italy}

\textbf{Author ORCID}{K. Modin, https://orcid.org/0000-0001-6900-1122; M. Viviani, https://orcid.org/0000-0002-2340-0483}

\textbf{Declaration of interests}{The authors report no conflict of interest.}

\section*{Appendix A: Stream function splitting}\label{sec:stream_func}
In this section, we present the analogous splitting of Section~\ref{sec:vort_split} for the stream function $P$.
Let $\Stab_W$ be the stabilizer of $W$ in $\SU(N)$. 
Then, we can consider the stream function splitting $P = P_s + P_r$, where $P_s$ is the orthogonal projection of $P$ onto $\Stab_W$, and $P_r$ the orthogonal complement. 
Similarly to what we have done in Section~\ref{sec:vort_split}, $P_s$ can be constructed by finding $U$ unitary which diagonalizes $W$:
\[
U^*WU=D,
\]
and then defining $P_s=U\mbox{diag}(U^*WU)U^*$. By definition the quantized Euler equations for the stream function $P$ can be written as:
\begin{equation*}
\dot{P} =\Delta_N^{-1}[P_r,\Delta(P_s+P_r)],
\end{equation*}
the energy $H(P) = \frac{1}{2}\Tr(\Delta_N(P_s+P_r)P_s)=\frac{1}{2}\Tr(\Delta_N(P_s)P_s)-\frac{1}{2}\Tr(\Delta_N(P_r)P_r)$ and the enstrophy $E(P) = -\Tr((\Delta_N(P_s+P_r))^2)$.
Notice that the enstrophy $E(P) = -\Tr((\Delta_N P)^2)$, in the $P_s,P_s$ does not admit a splitting contrary to the energy $H(P) = \Tr(\Delta_N P P)$:
\begin{align*}
H(P) &= H(P_s) - H(P_r)\\
E(P) &\neq E(P_s) + E(P_r).
\end{align*}
Since the eigenvalues of $W$ are constant in time, the dynamics of $P_s,P_r$ variables is simpler than the one of $W_s,W_r$. 
In fact, we have that $\dot{U}U=P$ and so 
\begin{align*}
\dot{P}_s &= [\dot{U}U,P_s] - \Pi_W[\dot{U}U,P_r] + \Pi_W(\dot{P})\\
&=[P_r,P_s] + \Pi_W(\Delta_N^{-1}[P_r,\Delta_N(P_s+P_r)]),
\end{align*}
and
\begin{align*}
\dot{P}_r &= -[P_r,P_s] + \Pi_W^\bot(\Delta_N^{-1}[P_r,\Delta_N(P_s+P_r)]).
\end{align*}
To understand how the dynamics of the $P_s,P_r$ variables look, we consider the first numerical simulation of Section~\ref{sec:vort_numerics}.
\begin{figure*}
\centering
\subfloat[Stream function field $P(t_{0})$]{
  \includegraphics[width=.45\textwidth]{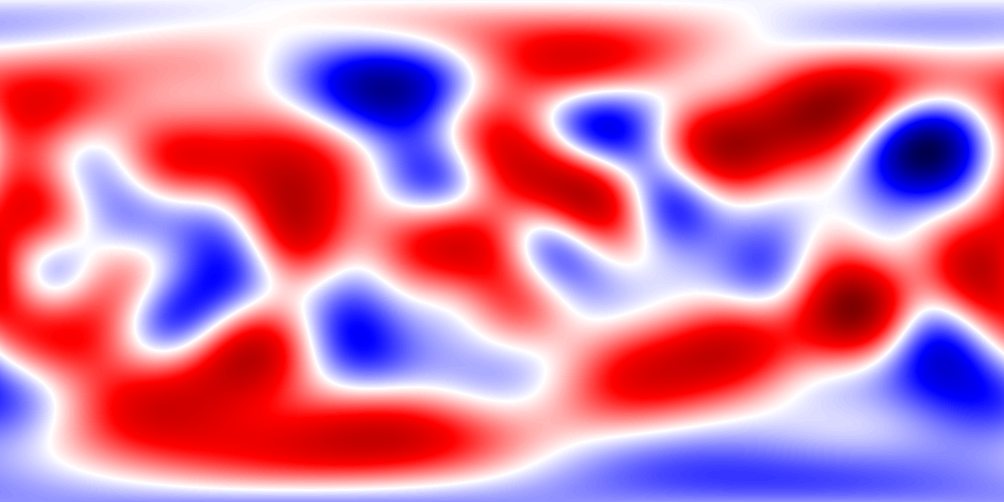}}
\subfloat[Stream function field $P(t_{end})$]{
  \includegraphics[width=.45\textwidth]{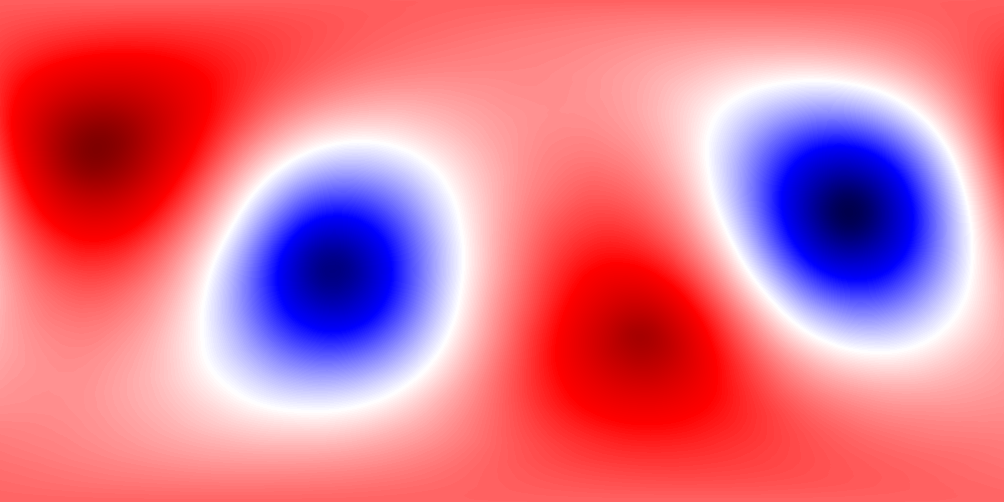}}\\
  \subfloat[Stream function field $P_s(t_{0})$]{
  \includegraphics[width=.45\textwidth]{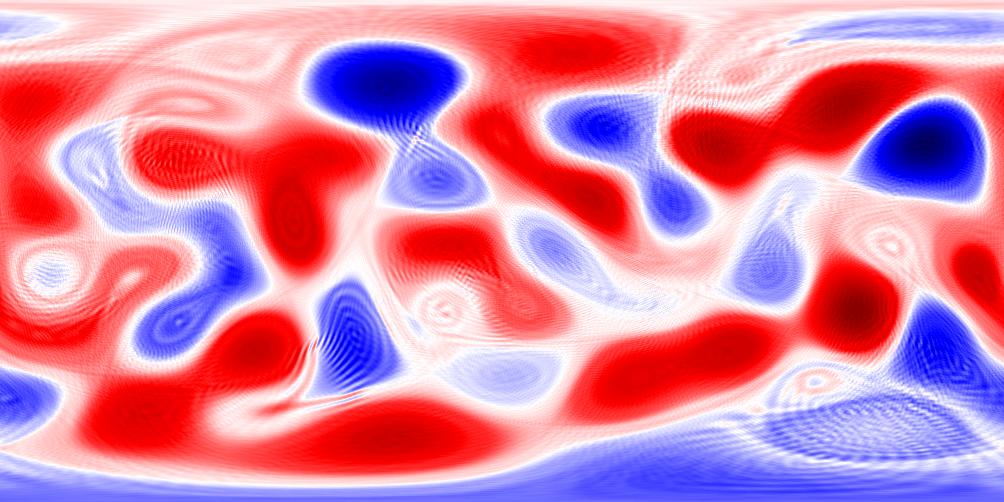}}
\subfloat[Stream function field $P_s(t_{end})$]{
  \includegraphics[width=.45\textwidth]{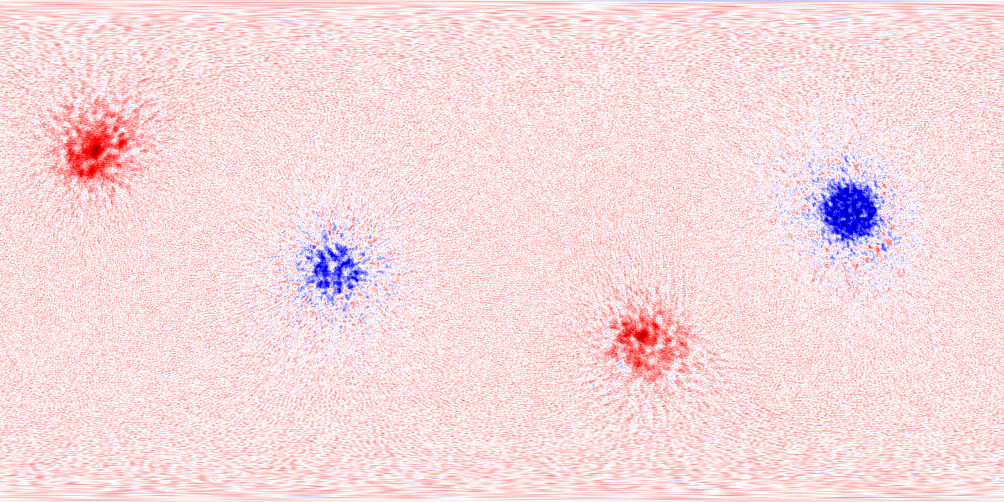}}\\
  \subfloat[Stream function field $P_r(t_{0})$]{
  \includegraphics[width=.45\textwidth]{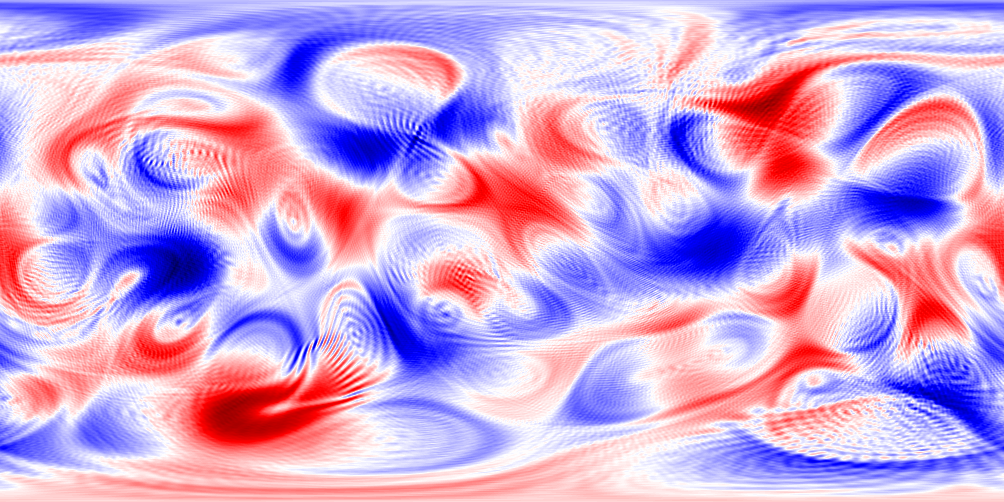}}
\subfloat[Stream function field $P_r(t_{end})$]{
  \includegraphics[width=.45\textwidth]{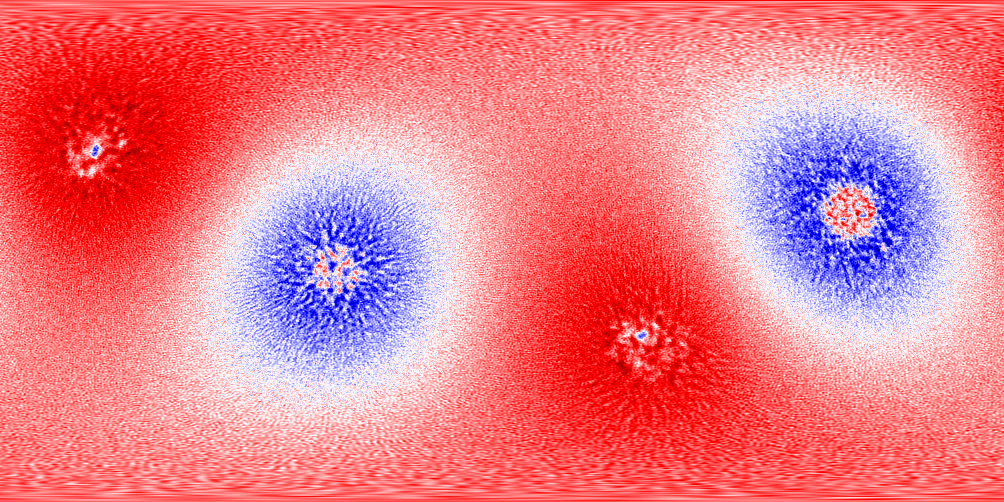}}
\caption{The Stream function fields fields $P,P_s,P_r$ at $t=0$ and $t=t_{end}$.}\label{fig:Stream_function}
\end{figure*}
In Figure~\ref{fig:Stream_function} the three stream function fields $P,P_s,P_r$ are shown at the times $t=0$ and $t=t_{end}$.
We notice that from a very smooth $P$, the projection $\Pi_W$ onto the stabilizer rougher field $W$ produce a much coarser image.
In particular, both $P_s$ and $P_r$ do not show any additional structure or scale separation unlike to $W_s$ and $W_r$.

\bibliographystyle{plainnat}
\bibliography{biblio}

\end{document}